\documentclass[%
reprint,
amsmath,amssymb,
aps,
floatfix,
]{revtex4-1}

\usepackage{graphicx}
\usepackage{dcolumn}
\usepackage{amsmath}
\usepackage{amssymb}
\usepackage{bm}
\usepackage[bbgreekl]{mathbbol}
\usepackage{color} 
\usepackage{braket}
\usepackage{gensymb}
\usepackage[dvipsnames]{xcolor}
\usepackage{lpic}

\usepackage[percent]{overpic}
\usepackage{stackengine}
\usepackage{physics}
\usepackage{siunitx}
\usepackage{textcomp}
\usepackage{slashbox}
\usepackage{lipsum}
\usepackage{xr}
 \usepackage{hyperref}
 \hypersetup{
    colorlinks=true,
    linkcolor=blue,
    filecolor=magenta,
    citecolor=blue,
    urlcolor=cyan,
}
\usepackage[mathscr]{eucal}

\usepackage{amsfonts}
\usepackage{caption}
\usepackage{subcaption}
\usepackage[colorinlistoftodos]{todonotes}
\usepackage{mathtools}
\usepackage{empheq}
\usepackage{esvect}
\usepackage{rotating}

\newcommand{\MM}[2][black]{\textcolor{#1}{#2}}
\newcommand{\HM}[2][black]{\textcolor{#1}{#2}}

\newcommand{\q}{\textup{$\hat{q}$}}
\newcommand{\p}{\textup{$\hat{p}$}}

\DeclareSymbolFontAlphabet{\mathbbm}{bbold}
\DeclareSymbolFontAlphabet{\mathbb}{AMSb}

\begin{document}

\preprint{APS/123-QED}


\title{Hidden entanglement in twin beams generated through optical parametric amplification in hot alkali atoms}

\author{R. L. Rinc\'on Celis$^{1,2}$}
\email{raullrinconc@gmail.com}
\author{G. Nirala$^3$}
\author{A. Monta\~na Guerrero$^1$}
\author{T. L. Meireles$^1$}
\author{M. Martinelli$^1$}
\author{A. M. Marino$^{3,4}$}
\author{H. M. Florez$^1$}
\email{hans@if.usp.br}

\affiliation{$^1$ Instituto de F\'{\i}sica, Universidade de S\~ao Paulo, 05315-970 S\~ao Paulo, SP-Brazil}
\affiliation{$^2$ Laboratoire Kastler Brossel, UPMC-Sorbonne Université: 4 place Jussieu F-75252 Paris, France.}
\affiliation{$^3$ Center for Quantum Research and Technology and Homer L. Dodge Department of Physics and Astronomy, The University of Oklahoma, Norman, Oklahoma 73019, USA}
\affiliation{$^4$ Quantum Information Science Section, Computational Sciences and Engineering Division,
Oak Ridge National Laboratory, Oak Ridge, TN, 37831, USA} \thanks{This manuscript has been authored in part by UT-Battelle, LLC, under contract DE-AC05-00OR22725 with the US Department of Energy (DOE). The publisher acknowledges the US government license to provide public access under the DOE Public Access Plan (\texttt{http://energy.gov/downloads/doe-public-access-plan}).}

\date{\today}

\begin{abstract}
Proper characterization of quantum correlations in a multimode optical state is critical for applications in quantum information science; however, the most common entanglement measurements can lead to an incomplete state reconstruction. This is the case for the ubiquitous spectral measurement of field quadratures for which a full characterization of the quantum correlations between optical \MM{beams} is not possible. We demonstrate this effect in twin beams generated through parametric amplification by four-wave mixing in hot rubidium vapor, showing the role of a frequency dependent gain response. We implement a resonator-based measurement that reveals entanglement between \MM{beams} that is otherwise hidden by usual spectral measurements.
Additionally, this system shows how the phase shifts between the carrier and the sidebands on the involved fields affect the observation of entanglement for different entanglement witnesses, demonstrating the relevance of making a complete state tomography. 


\end{abstract}

\pacs{Valid PACS appear here}

\maketitle


Quantum correlated light fields are fundamental resources for developing a quantum network for applications in quantum communication and quantum computation \cite{Nielsen11}.
Optimal use of these quantum resources relies on the complete reconstruction of the quantum state.
In the continuous variable (CV) domain, state reconstruction is based on measurements of field quadratures of distinct modes of the field. In this case, typical measurements involve the spectral analysis of the photocurrent generated by either direct intensity or homodyne measurements of the field.

The photocurrent spectrum provides information about the beatnotes between the central frequency of a bright field, which can be treated as a carrier during the detection, and neighboring sidebands, which correspond to the different modes of the field. Typically, the noise analysis in this photocurrent cannot distinguish the contribution of upper and lower sidebands, thus leading to an incomplete information of the state. However, the ability to manipulate the sidebands prior to detection can serve as a tool for complete state reconstruction of the sideband modes~\cite{Assumption2020}.

We present a case under which entanglement involving a pair of beams is not observed, unless we make a detailed analysis of their sidebands. Moreover, even if the beams show entanglement, a stronger signature of nonclassicality can be obtained between a specific pair of modes. Such a complete characterization of the quantum correlations between sideband modes makes it possible to optimize available  quantum resources when performing a given quantum operation involving those modes.

Among the distinct tools to generate entangled states, optical parametric amplifiers (OPAs) utilizing a four-wave mixing (4WM) process in  alkali atomic vapors offer interesting possibilities to entangle multiple fields, ranging from two-mode squeezed states \cite{mccormick2007strong,glorieux2010double}, spatially multiplexed entangled states \cite{zhang2020reconfigurable,dong2022generation}, with applications in quantum communication, sensing and imaging \cite{boyer2008entangled,defienne2024advances}, among others. An interesting feature of these amplifiers is their relatively narrow bandwidth gain, as compared to their solid state counterparts. Such a narrow bandwidth, of the order of tens of MHz, may result in different amplification factors for each sideband, thus offering the possibility of generating twin beams with a hidden entanglement structure that cannot be measured using the traditional spectrally broadband media.

In this paper, we show the presence of a hidden quantum-correlated structure in the twin beams generated via a 4WM process in an isotopically pure $^{85}$Rb vapor cell. \MM{We perform a full tomography by individually addressing each sideband mode~\cite{Assumption2020}. In order to achieve this, we use the dispersive response of an auxiliary resonator as part of the detection system by reflecting a beam from the resonator before detection with a photodiode.} Such a resonator-assisted detection on each of the twin beams can achieve a complete reconstruction of the two-beam covariance matrix. Therefore, we are able to measure the asymmetry in the quantum correlations between the sideband modes of the twin beams generated using the 4WM process. 
We support our experimental observations through a theoretical study of the correlations in the twins beams based on the microscopic model of the 4WM used in Ref. \cite{glorieux2010double}, predicting  to some degree the asymmetry of the quantum correlations between sideband pairs.

\begin{figure}[h]
\centering
\captionsetup{justification=raggedright, width=0.46\textwidth}
	\begin{subfigure}{0.16\textwidth}
	\centering
    \includegraphics[width=0.98\textwidth]{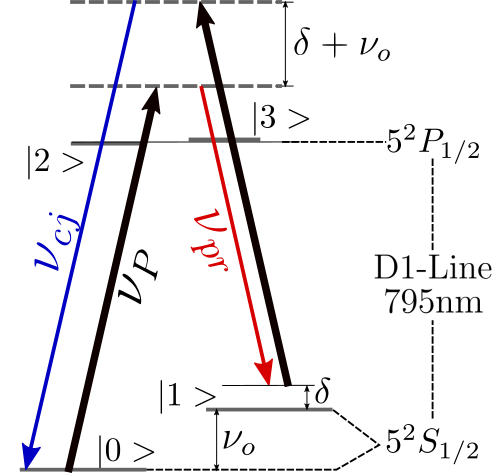}
    \caption{}
    \label{fig:FWM_Double-Lambda}
    \end{subfigure}
    \begin{subfigure}{0.31\textwidth}	
    \centering
    \includegraphics[width=0.95\textwidth]{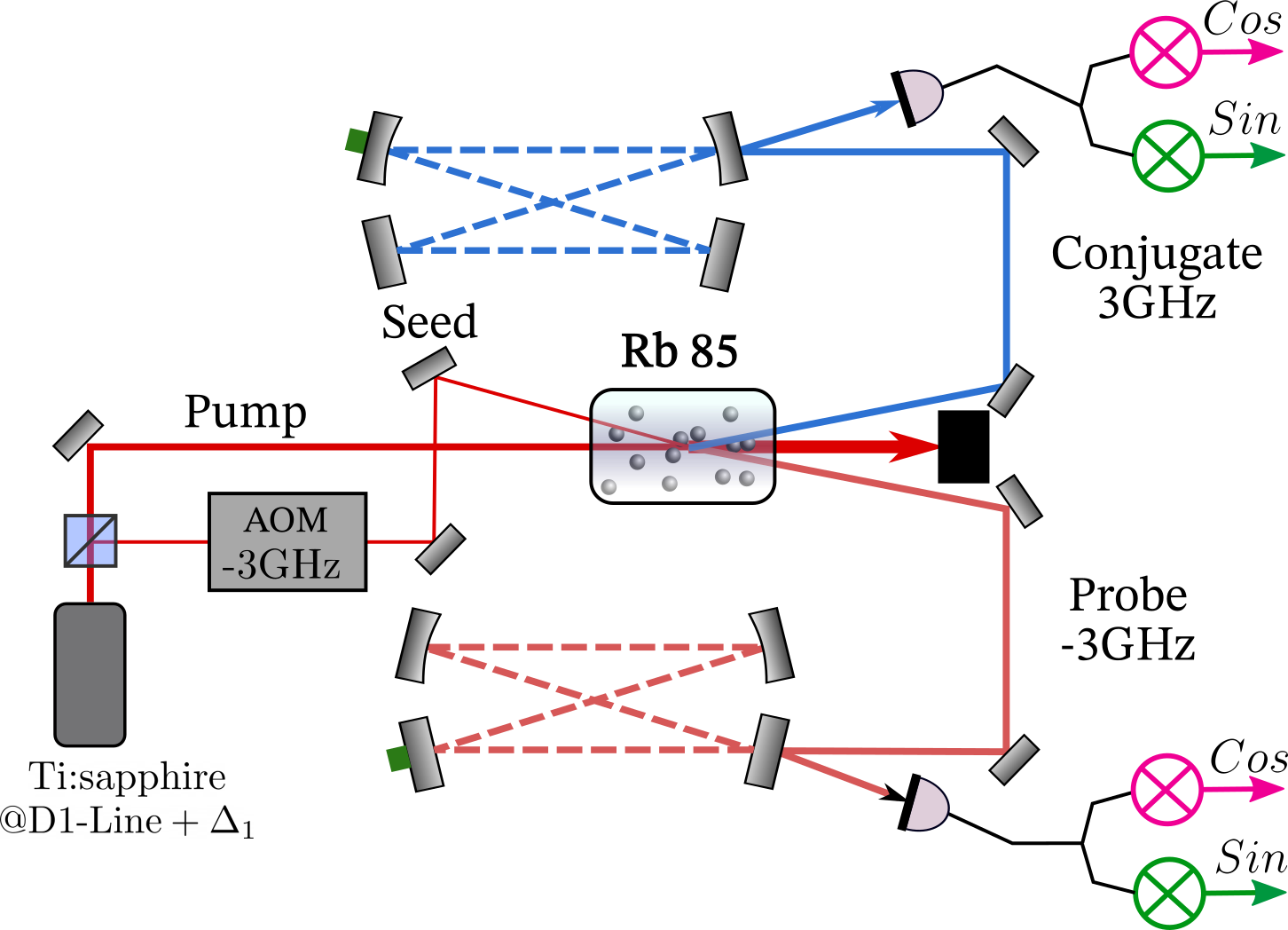}
    \caption{}
    \label{fig:Exp_setup}
    \end{subfigure}
    \caption[]{ \textbf{(a)} Double-lambda energy level structure on the D1 line of $^{85}$Rb used to implement the 4WM process. \textbf{(b)} Schematic of experimental setup.\label{fig:4WM}}
\end{figure}

The 4WM is a parametric process in which two fields, the probe (or seed) and the pump, interact with a third order ($\chi ^{3}$) non-linear medium to amplify the probe and generate a new field, the conjugate. \autoref{fig:4WM}\textcolor{blue}{(a)} shows the double-lambda energy level structure used for the the 4WM process in the D1 line of $^{85}$Rb ($\sim$795 nm). The pump beam (black bold arrows, frequency $\nu_P$) couples ground states $\ket{0}$ and $\ket{1}$ (associated to hyperfine levels $F=2$ and $F=3$, which are separated by a frequency $\nu_{0}\sim$3 GHz) to virtual excited states (dashed horizontal lines). 
The probe beam (red arrow, frequency $\nu_{pr}$) is red detuned by $\sim$3 GHz from the pump frequency, and nearly closes a lambda-transition with a two-photon detuning $\delta=\nu_{P}-\nu_{pr}-\nu_{0}$. 
\MM{The probe acts as a seed that is amplified by the 4WM process, while stimulating the generation of the conjugate beam }(blue arrow, frequency $\nu_{cj}$), blue detuned by $\sim$3 GHz from the pump frequency. 

\HM{This amplification through a 4WW} process can be described as quantized fields interacting with a four level system, based on the microscopic model used in Ref. \cite{glorieux2010double}. \HM{Such model considers the probe and the conjugate fields, represented by $\hat{a}$ and $\hat{b}$ respectively, interacting} with the pump field via the Rb atoms while propagating through the medium. 
\HM{This interaction leads to amplification which can be simply described as the propagation of the vector  $ \hat{\mathbf{A}}(z)=(\hat{a},\hat{a}^\dagger,\hat{b},\hat{b}^\dagger)$  such that }
\begin{align}
 \hat{\mathbf{A}}(z)=&e^{\mathbf{R} z} \hat{\mathbf{A}}(0),
 \label{eq:field}
\end{align}
\HM{in which the initial fields  $\hat{\mathbf{A}}(0)$ are amplified by a factor $e^{\mathbf{R} z}$ along a distance $z$ where the matrix $\mathbf{R}$ includes all the parameters of the interaction, e.g. optical detuning, Rabi frequencies, and spontaneous decay rate. Section I of the supplementary material shows a more detailed description of eq.(\ref{eq:field}).}
\HM{From this amplitude we can determine the intensity of the output fields as well as the noise properties of the field, which we can later compare to the experimental results. In the case of the intensity, the output fields is}  given by $I= \mathcal{J}(z) \langle \hat{\mathbf{A}}(0)\hat{\mathbf{A}}(0)^T \rangle \mathcal{J}(z)^T $, where the propagator is given by $\mathcal{J}(z)=e^{\mathbf{R}z}$. 
\HM{In the case of the noise properties, by}
linearizing the field operators,  we can write $\hat{\mathbf{A}}(z)=\langle \hat{\mathbf{A}}(z) \rangle  + \delta \hat{\mathbf{A}}(z)$, where $ \delta \hat{\mathbf{A}}(z)$ corresponds to the field fluctuations. The evolution of the field fluctuations after propagating through the medium is similar to that of  Eq. (\ref{eq:field}), but in the frequency domain,  
\begin{align}
 \delta \hat{\mathbf{A}}(\omega,z)=&e^{\mathbf{R}(\omega) z}  \delta \hat{\mathbf{A}}(\omega,0),
 \label{eq:d_field}
\end{align}
where $\mathbf{R}(\omega)$ has the same form as $\mathbf{R}$  but is frequency dependent, as shown in Sec. II of the Supplementary Material. From the field fluctuations \HM{$ \delta \hat{\mathbf{A}}(\omega,z)$} we can reconstruct the state of the probe and conjugate beams through the covariance matrix.



\HM{Our experiment setup is shown in} Fig.~\ref{fig:4WM}\textcolor{blue}{(b)} \HM{ in which we utilize}  a continuous wavelength (CW) stabilized Ti:sapphire laser, blue-shifted from the $5^2S_{1/2},F=2\rightarrow 5^2P_{1/2},F=2$ transition of $^{85}$Rb by $850$ MHz. Most of the power of the laser (nearly 400~mW) is used as the  pump for the 4WM process, while a small fraction passes through an acousto-optic-modulator to red-shift its frequency by $\sim $3 GHz and produce a $\sim $100 $\mu$W seed probe beam. Both the probe seed and the pump beams, with a waist of 244 $\mu$m and 525 $\mu$m respectively, are directed into the vapor cell  maintained at a temperature of 103$^\circ $C. After the 4WM, the amplified probe and conjugate beams are reflected from cavities before detection with a photodiode to implement the resonator-assisted detection for the covariance matrix reconstruction, following the technique used in Ref. \cite{Assumption2020} and also detailed in the Supplementary Material.

\begin{figure}[h!]
    \centering
    \captionsetup{justification=raggedright, width=0.46\textwidth}
    \includegraphics[width=0.48\textwidth]{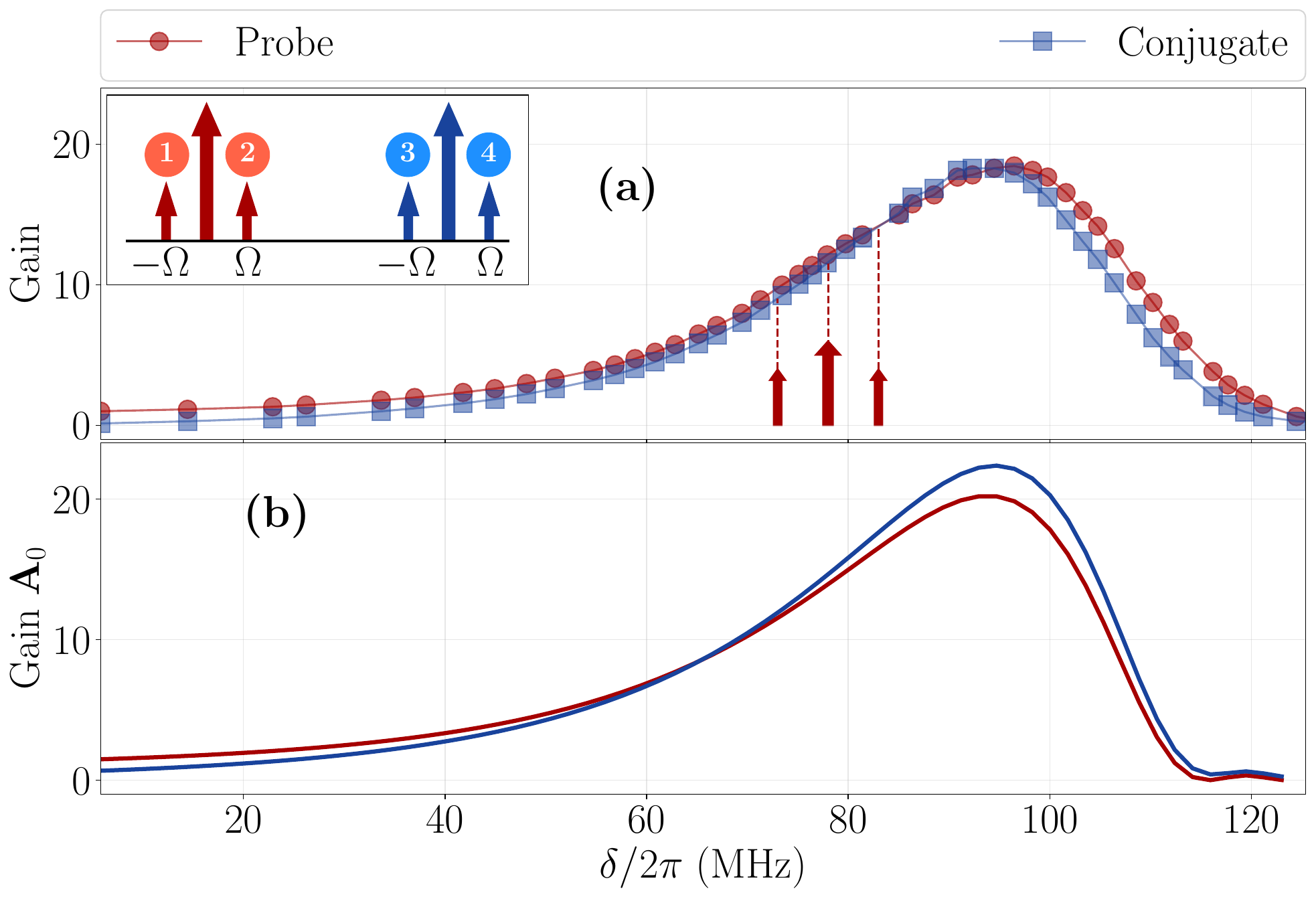}
    \caption{\textbf{(a)} Experimental gain profile. Inset: representation of the twin beams in the frequency domain represented as a central carrier with upper and lower sidebands for each of the beams. Numbers 1 and 3 corresponds to the lower sidebands of the probe/conjugate; while 2 and 4 corresponds to the upper sidebands. \textbf{(b)} Theoretical gain profile for the probe (red) and conjugate (blue).
    }
    \label{fig:Gain_amp}
\end{figure}

The measured 4WM amplification gain spectrum of the probe/conjugate beam is shown in \autoref{fig:Gain_amp}\textcolor{blue}{(a)}.
Notice that for a given frequency of the probe, its upper and the lower sidebands components (associated with $\pm \Omega$, see inset in Fig. \ref{fig:Gain_amp}\textcolor{blue}{(a)}) are amplified by different factors. This imbalance is enhanced as $\delta $ gets closer to the maximum gain. The gain profile determines the strength of the quantum correlations between the sidebands. Figure~\ref{fig:Gain_amp}\textcolor{blue}{(b)} presents the gain profile from the theoretical model, which shows an accurate description for the amplification of both fields. 

The field observables can be represented by the quadratures associated to the amplitude $\hat{p}_{\pm\Omega}$ and phase $\hat{q}_{\pm\Omega}$
of the upper and lower sidebands, labeled as $\pm\Omega$. 
These operators can be expressed in terms of the creation and annihilation operators as $\hat{a}_{\pm\Omega}=(\hat{p}_{\pm\Omega}\pm  i\hat{q}_{\pm\Omega})$, and  satisfy the commutation relations $[\hat{p}_{\pm\Omega},\hat{q}_{\pm\Omega'}]=2i\delta(\Omega-\Omega')$ and $[\hat{p}_{\pm\Omega},\hat{q}_{\mp\Omega'}]=0$. 
\MM{However, when directly detected with photodiodes, either for intensity or homodyne measurements, the measured quadratures are associated with symmetric/anti-symmetric (SA) combinations of these quadratures, which can be written as \cite{Assumption2020}}
\begin{eqnarray}
\label{eq:psa_psd}
\p_{S}=\left(\p_{+\Omega}+ \p_{-\Omega}\right)/\sqrt{2}, \hspace{0.2cm}\p_{A}=\left(\p_{+\Omega}- \p_{-\Omega}\right)/\sqrt{2},\\
\label{eq:qsa_qsd}
\q_{S}=\left(\q_{+\Omega}+ \q_{-\Omega}\right)/\sqrt{2}, \hspace{0.2cm}\q_{A}=\left(\q_{+\Omega}- \q_{-\Omega}\right)/\sqrt{2}.
\end{eqnarray}

The state of the twin beams produced by the 4WM process can be assumed to be a Gaussian state, which aside from a displacement, is completely characterized by the covariance matrix. In the SA basis $\hat{X}_{SA}$, the covariance matrix takes the form:

\begin{eqnarray}
\nonumber \mathbb{V}_{SA}&=&\frac{1}{2}\left( \langle \hat{X}_{SA}\cdot \hat{X}^T_{SA}\rangle + 
    \langle \hat{X}_{SA}\cdot \hat{X}^T_{SA}\rangle^T \right),\\
\mathbb{V}_{SA} &=& \begin{pmatrix}
\mathbb{V}_S & \mathbb{C}_{SA}\\
\mathbb{C}_{SA} & \mathbb{V}_A
\end{pmatrix},
\label{eq:Covariance_sa_block}
\end{eqnarray}
where $\hat{X}_{SA}=\left(\hat{X}_S,\hat{X}_A\right)^T$ is a column vector composed of the symmetric/anti-symmetric quadratures $\hat{X}_{j} = \left(\p_{j}^{(pr)},\q_{j}^{(pr)},\p_{j}^{(cj)},\q_{j}^{(cj)}\right)$, with the subscript $j$=S,A. 
The covariance matrix in the SA basis is organized in blocks. The diagonals correspond to the symmetric ($\mathbb{V}_S$) or anti-symmetric parts ($\mathbb{V}_A$), and the off-diagonal terms $\mathbb{C}_{SA}$ to the cross-correlations between the quadratures of different modes (check Supplementary Material).

\MM{In order to express it in terms of the sideband basis, we can reorganize Eqs. (\ref{eq:psa_psd}) and (\ref{eq:qsa_qsd}) in a matrix form such that the quadratures in the SA and the sideband basis can be related through the linear transformation $\Lambda$ such that $\hat{X}_{\pm\Omega}=\Lambda \hat{X}_{SA}$. Therefore, the covariance matrix follows:}
\begin{equation}
    \mathbb{V}_{\pm\Omega} = \Lambda \mathbb{V}_{SA}\Lambda ^T.
\end{equation}

Note that in the sideband basis, the terms of the form $\langle \p_{\pm\Omega}^{(j)}\,\q_{\pm\Omega}^{(j)}\rangle$ are zero, and $\Delta ^2 \p_{\pm\Omega}=\Delta ^2 \q_{\pm\Omega}>1$. Therefore, the modes in the sideband basis correspond necessarily to thermal states, unlike in the SA basis \cite{Assumption2020}.

The elements of these covariance matrices can be fully characterized by employing the \emph{resonator-assisted detection} 
\cite{Martinelli:23,barbosa2013beyond,Assumption2020}. In the experimental setup, the resonators are arranged in a bow-tie configuration, comprised of two plane mirrors and two curved mirrors with a focal length of $1$~m. Their bandwidth is $\Gamma/2\pi = (3.37\pm0.01)$~MHz, with a finesse of $(87.4\pm0.2)$, and a power depletion of 50\% on resonance. To ensure a complete characterization of the state, measurements are done at an analysis frequency $\Omega/2\pi=7$~MHz, which satisfies the needed condition of $\Omega\geq \sqrt{2} \Gamma$ to obtain a proper rotation  of the sideband modes \cite{villar2008conversion}. 
The complete reconstruction requires 
the demodulation of the photocurrents using in-phase (Cos) and in-quadrature (Sin) reference oscillators,
as shown in Fig. \ref{fig:4WM}\textcolor{blue}{(b)}. 
\MM{The overall system detection efficiency is 83 \%, accounting for optical losses and the quantum efficiency of the photodetectors. The losses coming from the impedance matching with the cavity are considered in detail in the Supplementary Material}. The standard quantum level (SQL) is calibrated by replacing the probe and the conjugate with a coherent laser source of the same power.


\begin{figure}[hb]
    \centering
    \captionsetup{justification=raggedright, width=0.46\textwidth}
        \includegraphics[width=0.95\linewidth]{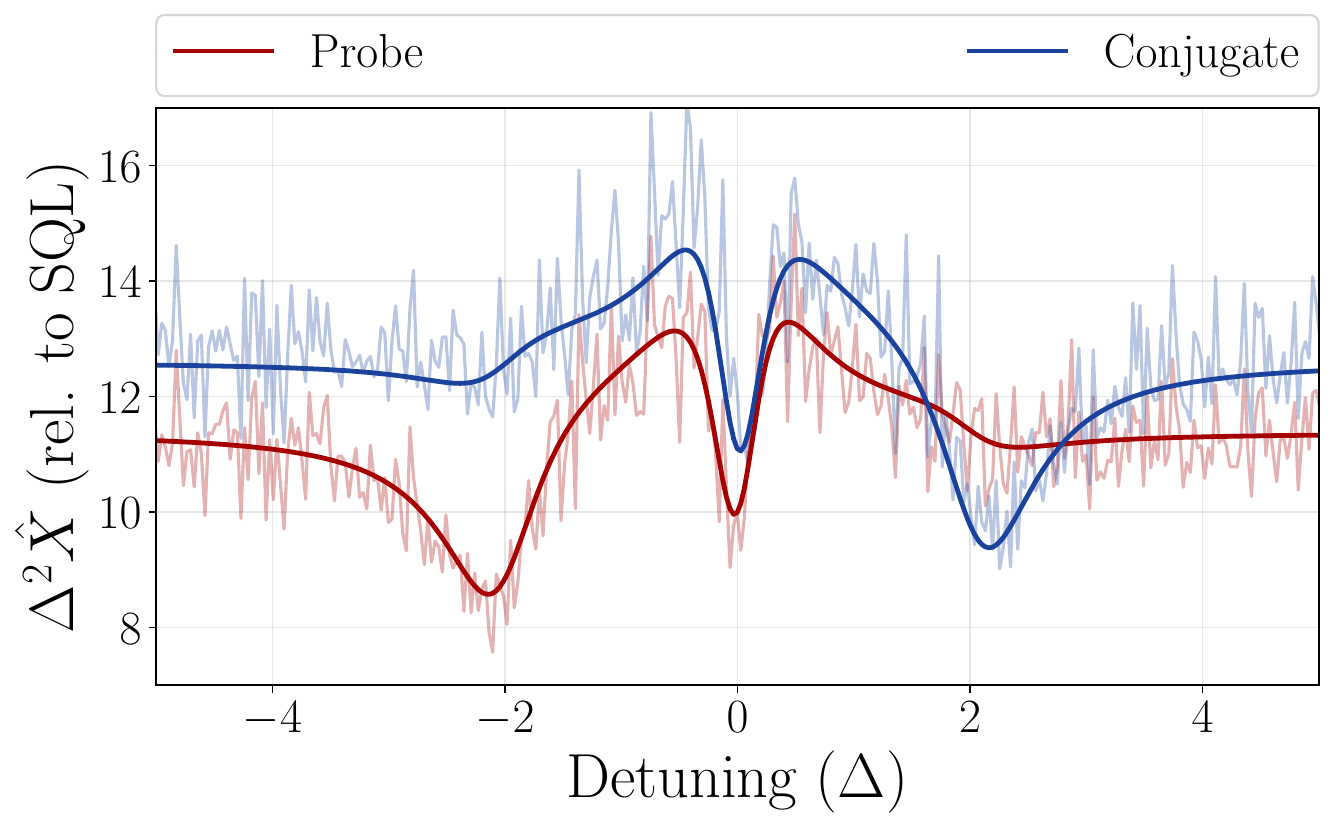}
        \caption{Individual noise spectral densities normalized to the corresponding SQL as a function of the normalized resonator detuning  ($\Delta$) for the probe (red) and conjugate (blue).  Probe detuning $\delta/2\pi \sim 75$ MHz. Analysis frequency $\Omega/2\pi=7$~MHz.}  
    \label{fig:Noise_spectral}
\end{figure}

\MM{\autoref{fig:Noise_spectral} displays the noise spectral density for the intensity measurement} after reflection from the analysis cavities for the probe (red) and the conjugate (blue) beams normalized to the corresponding SQL, as the resonance frequencies of the cavities are scanned around the corresponding carrier frequency. \MM{The acquisition time for the measurement is synchronized, but the scanning is done independently}. In this figure, the detunings are normalized to the cavity bandwidths. By tuning the cavity around $\Delta = -2$, as determined by the analysis frequency, the probe beam (red) exhibits a depletion dip that reaches $8.5$ SQL, while around $\Delta = 2$ the depletion is almost negligible, and the level is close to $11$ SQL. The conjugate (blue) beam exhibits a mirror behavior, with depletion around $\Delta = 2$.

Given that the curves show the depletion of the reflected field when its frequency is on resonance with the cavity, the strong depletion of the probe when its lower sideband is resonant with the cavity at $\Delta = -2$ demonstrates that most of the energy is in this mode. Due to the expected correlation between the sidebands as a result of energy conservation in the 4WM process, the same strong depletion for the conjugate occurs when its upper sideband is resonant, at $\Delta = 2$. This is consistent with the idea of the pairwise photon generation in the coupled modes in Fig.~\ref{fig:Gain_amp}\textcolor{blue}{(a)}.

The imbalance in energy between the upper and lower sidebands of the same beam is thus evident and can be evaluated by fitting the experimental data to Eqs. (\ref{supp-eq:Noise_spectral_density_single_field}) and (\ref{supp-eq:cross_corr_noise_spectrum}) of the Supplemetary Material. The results of the fit are given by the solid lines in Fig. \ref{fig:Noise_spectral}.
Given the average photon number of each sideband,  $\mathcal{E}_i(\pm \Omega )=(1/2)(\Delta ^2 \hat{q}_i(\pm \Omega) + \Delta ^2\hat{p}_i(\pm \Omega))-1$,
the energy imbalance (normalized to the photon energy $\hbar\omega_i$) can be directly evaluated as
\begin{equation}
    \Delta \mathcal{E}_i= \frac{1}{2}\left[\mathcal{E}_i(+ \Omega )-\mathcal{E}_i(- \Omega )\right].
\end{equation}
This term is also related to the correlation 
$\langle \hat{p}_s^{(i)}\,\hat{p}_a^{(i)} \rangle$ in the symmetric/anti-symmetric description \cite{Assumption2020}, which significantly impacts the measured entanglement properties when analyzed in the sideband basis, as will be discussed later.

\begin{figure}[h]
    \centering
    \captionsetup{justification=raggedright, width=0.46\textwidth}
    \begin{overpic}[width=0.49\linewidth]{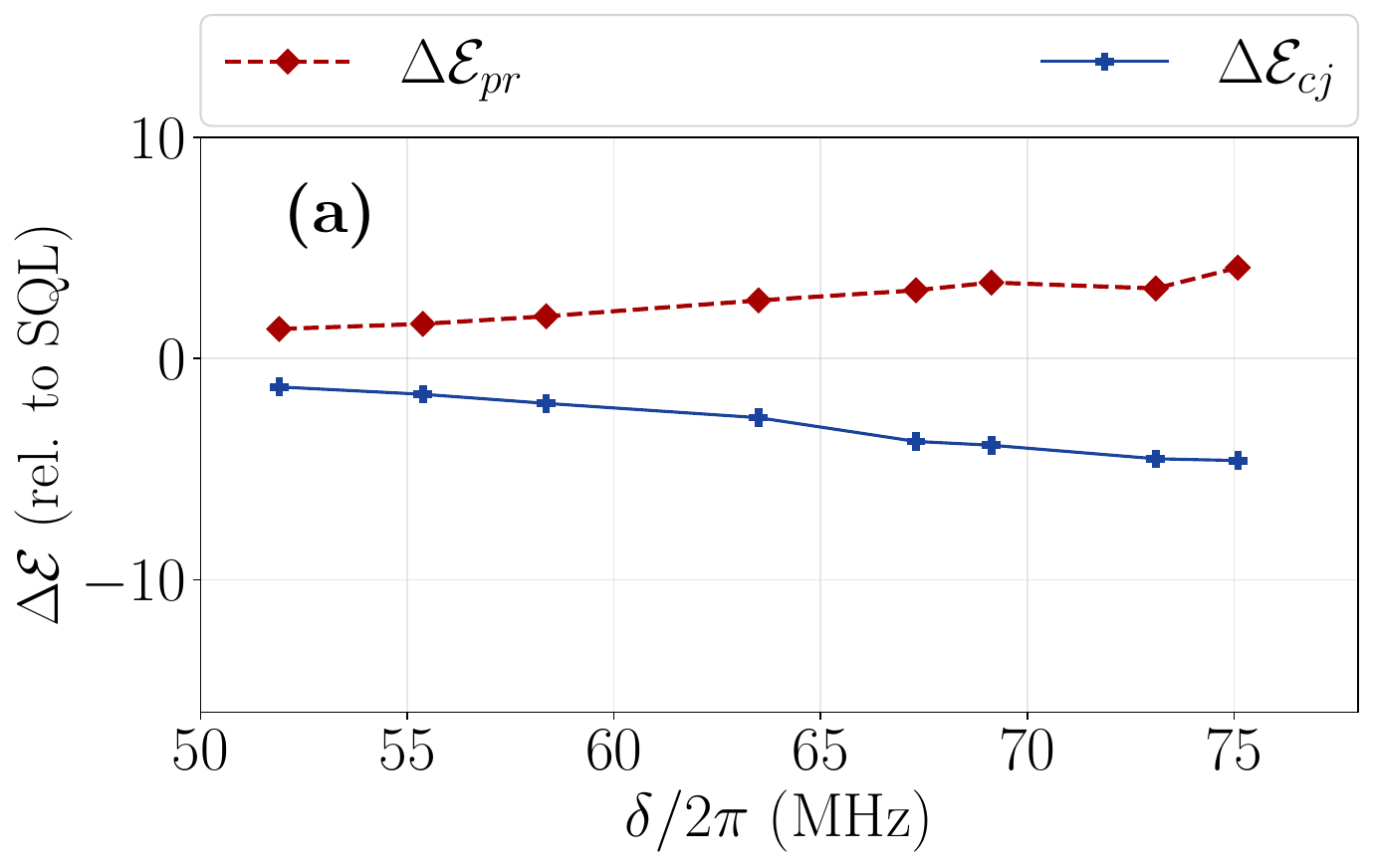}
    \end{overpic}%
    \begin{overpic}[width=0.49\linewidth]{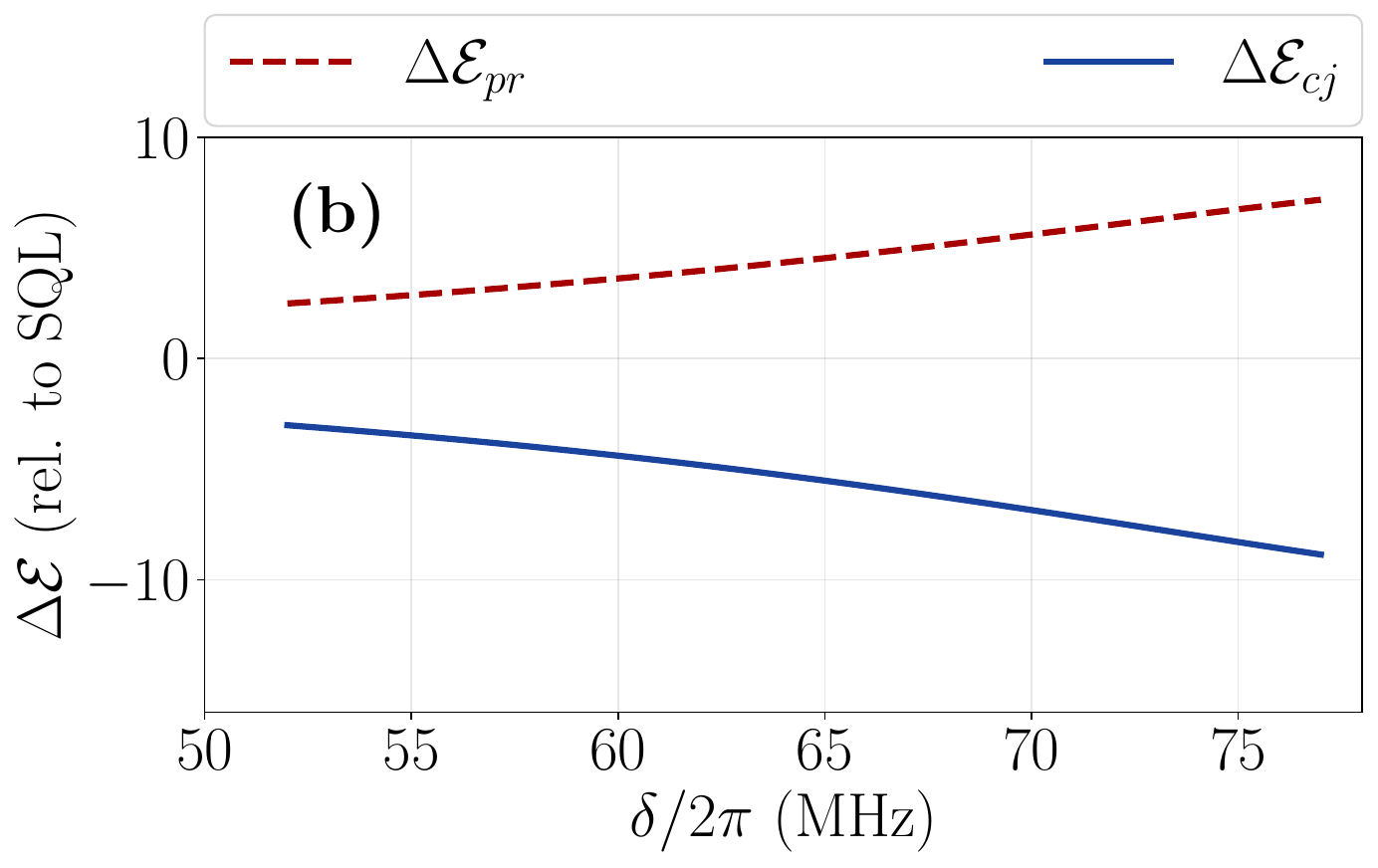}
    \end{overpic}
    \caption{\textbf{(a)} Experimental and \textbf{(b)} theoretical energy imbalance between the sidebands as a function of two-photon detuning for the probe (red) and conjugate (blue).}
    \vspace{-0.3cm} 
    \label{fig:delta_param}
\end{figure}

\MM{\autoref{fig:delta_param} depicts the energy imbalance $\Delta \mathcal{E}$ as a function of the two-photon detuning. 
\HM{As $\delta/2\pi$ increases from 50~MHz to 75~MHz, the energy imbalance increases as well. This is consistent with the fact that by increasing $\delta/2\pi \rightarrow 75$~MHz, the probe field is approaching the maximum gain of amplification which is at $\sim 95$MHz (see Fig. \ref{fig:Gain_amp}). }
The upper sideband for the probe ($\Omega_{pr}$, labeled 2 in the inset of Fig. \ref{fig:Gain_amp}\textcolor{blue}{(a)}) at $\sim 84$MHz experiences a gain factor of around 15 units, along with its corresponding twin, the lower sideband of the conjugate $-\Omega_{cj}$ (labeled 3). Conversely, the lower sideband of the probe, $-\Omega_{pr}$ (labeled 1), observed at $\sim 72$MHz, experiences a gain factor of around 9 units, along with its twin, the upper sideband of the conjugate $\Omega_{cj}$ (labeled 4).
 The developed model, shown in Fig. \ref{fig:delta_param}\textcolor{blue}{(b)}, gives a fair qualitative agreement with the experimental results shown in Fig. \ref{fig:delta_param}\textcolor{blue}{(a)}, which shows the observed increase in the magnitude of the energy imbalance as the probe beam approaches the maximum gain.}


A generally used entanglement test between two modes, known as the \HM{Duan, Giedke, Cirac, and Zoller (DGCZ)} criteria, involves the measurement of EPR-like operators \cite{duan2000inseparability}.
In our case, it involves $\hat{p} _{-} = (\hat{p}^{(1)} - \hat{p}^{(2)})/4$, and $\hat{q} _{+} = (\hat{q}^{(1)} + \hat{q}^{(2)})/4$ for two given modes (1,2), such that the separable state satisfies the inequality $\Delta ^2\hat{p}_-+\Delta ^2 \hat{q}_+\geqslant 1$. Violation of this condition is a sufficient inseparability condition, 
thus providing an entanglement witness.

\begin{figure}[h]
    \centering
    \captionsetup{justification=raggedright, width=0.46\textwidth}
    \begin{overpic}[width=0.49\linewidth]{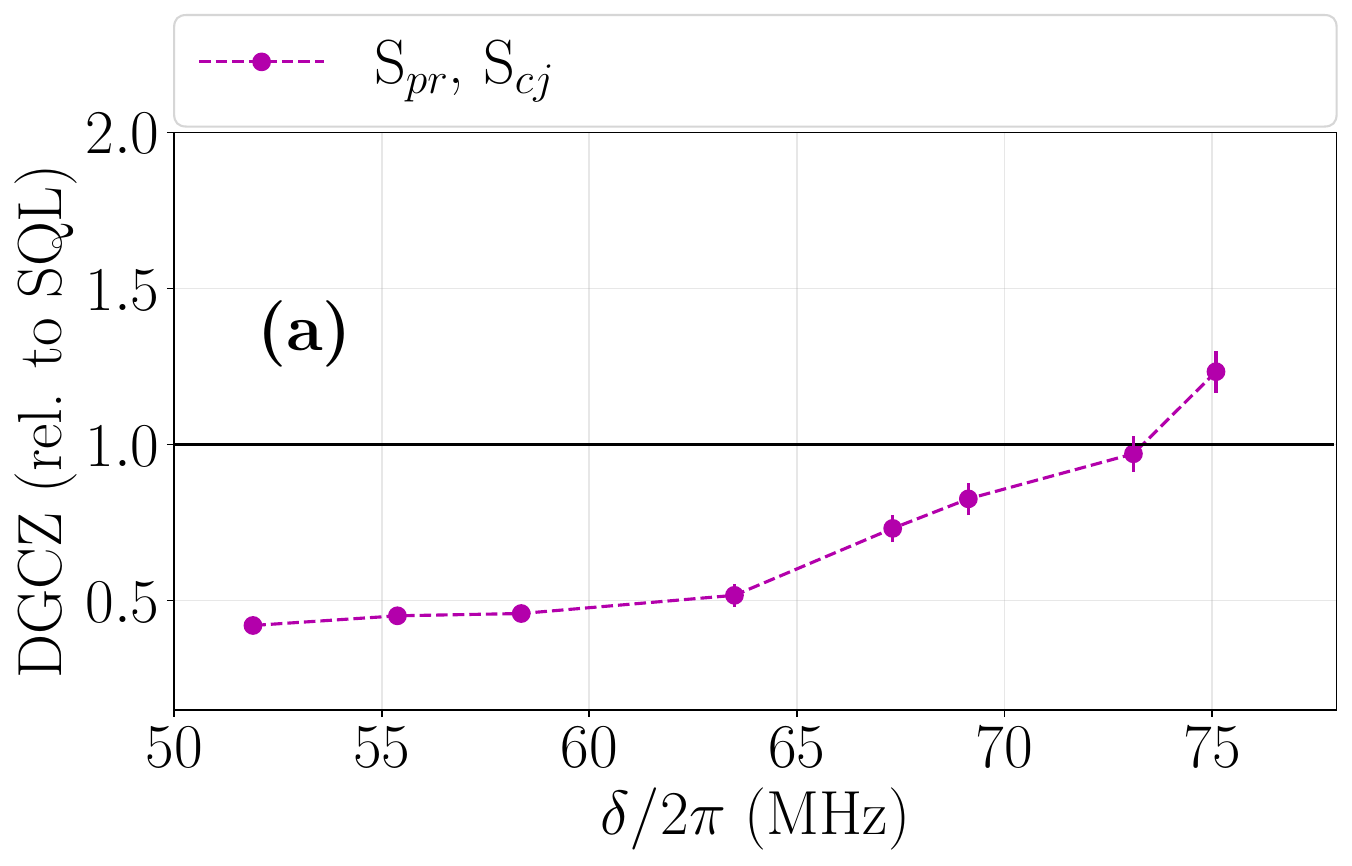}
    \end{overpic}%
    \begin{overpic}[width=0.49\linewidth]{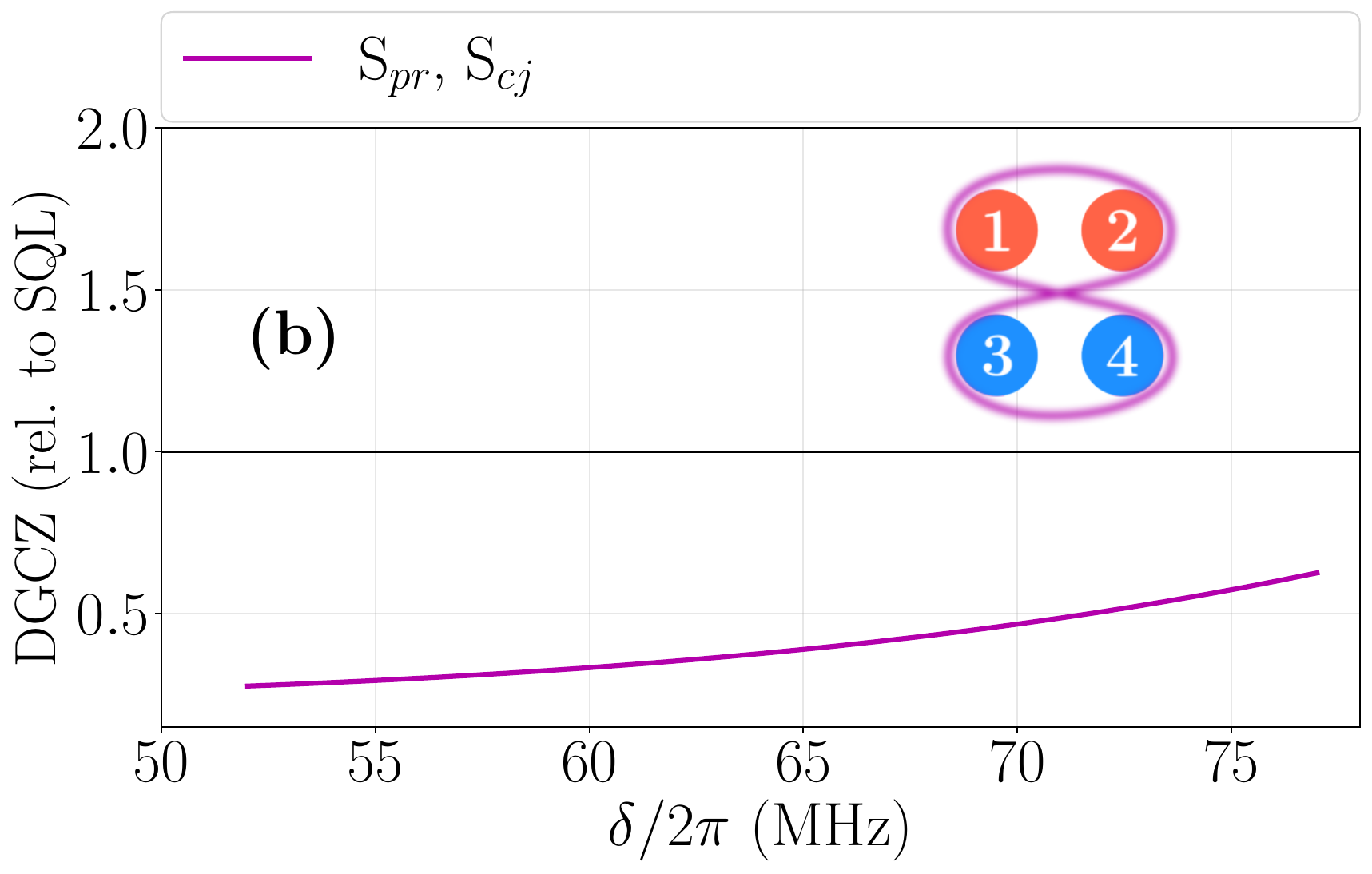}
    \end{overpic}
    \begin{overpic}[width=0.49\linewidth]{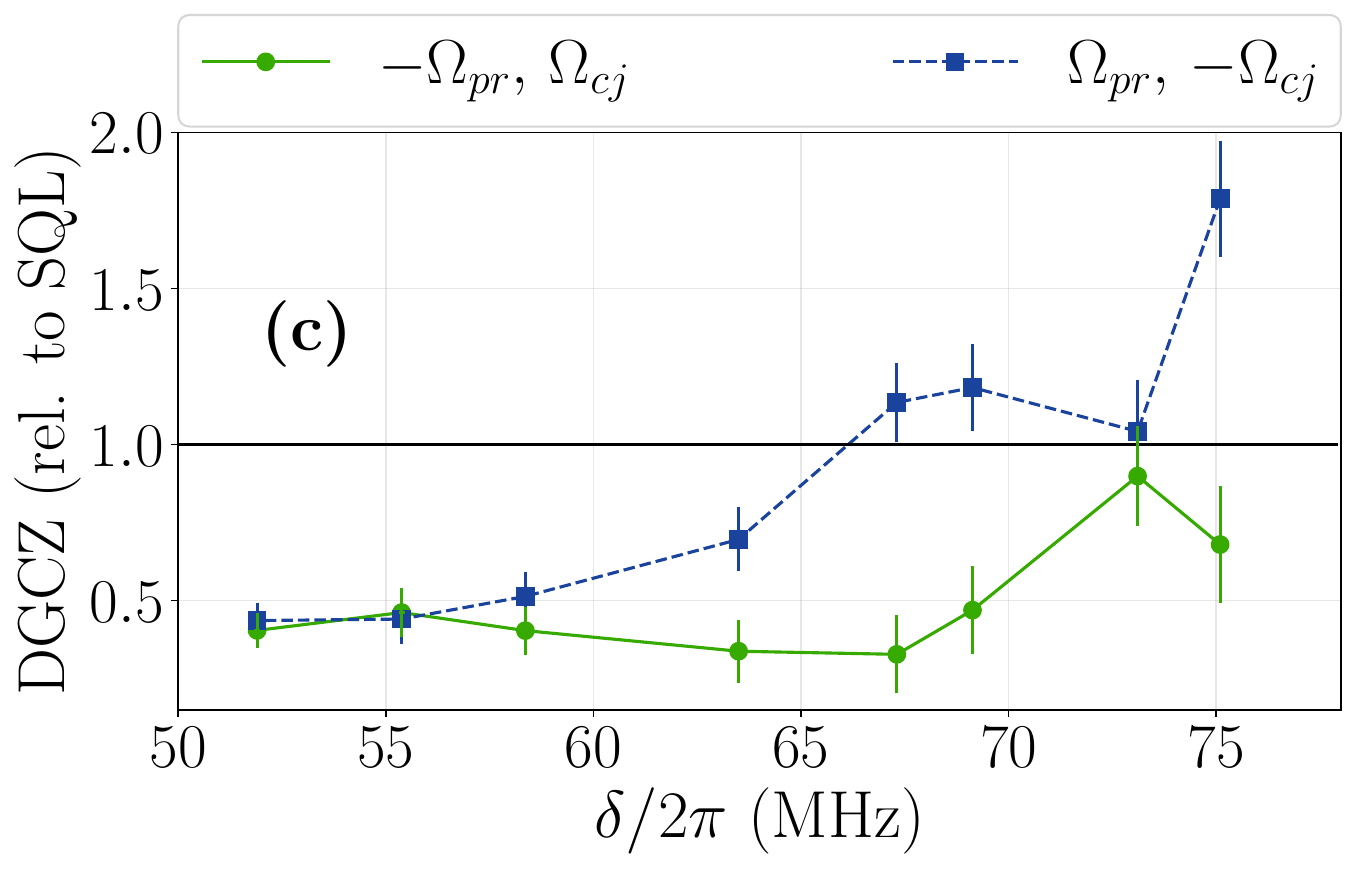}
    \end{overpic}%
    \begin{overpic}[width=0.49\linewidth]{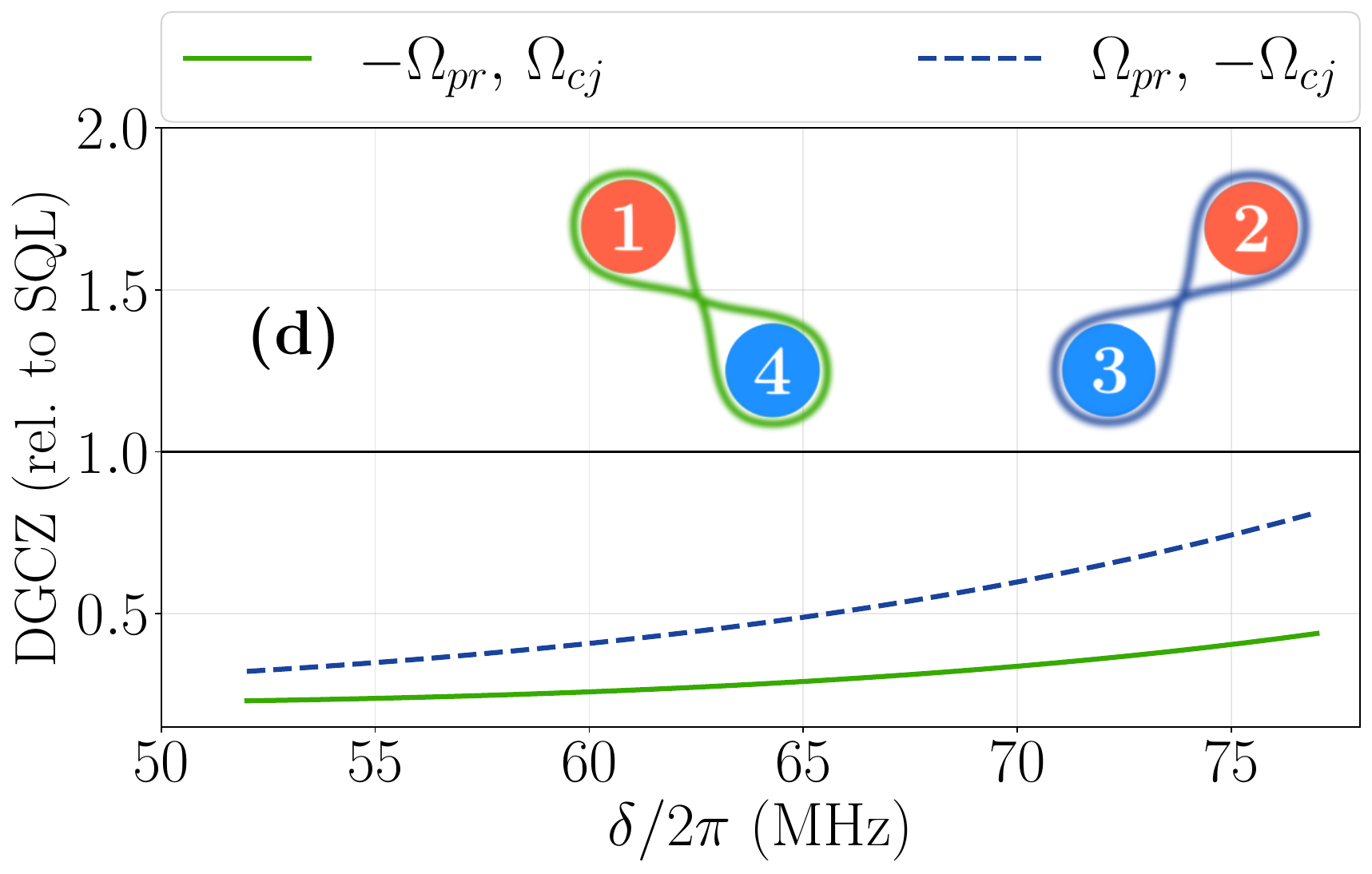}
    \end{overpic}
    \caption{Measured and theoretical DGCZ criteria in the SA basis, \textbf{(a)} and \textbf{(b)} respectively, and in the sideband basis, \textbf{(c)} and \textbf{(d)} respectively, as a function of two-photon detuning.}
    \label{fig:Duan}
\end{figure}

\autoref{fig:Duan} illustrates the experimental and theoretical outcomes for this criterion, presenting results in the SA basis in Figs. \ref{fig:Duan}\textcolor{blue}{(a)} and \ref{fig:Duan}\textcolor{blue}{(b)}, and for the relevant sideband pair combinations in the sideband basis in Figs. \ref{fig:Duan}\textcolor{blue}{(c)} and \ref{fig:Duan}\textcolor{blue}{(d)}. 
The figures show how the DGCZ entanglement witness changes as the two-photon detuning $\delta$ moves closer to the maximum of the gain profile.
The evaluation of quantum correlations in the SA basis is what is typically done with standard spectral measurements techniques \cite{mccormick2007strong,glorieux2010double,zhang2020reconfigurable,dong2022generation}. The label $\text{S}_{pr},\text{S}_{cj}$ \MM{refers to the DGCZ criteria evaluated for the symmetric bipartitions composed by the probe and conjugate beams. }
\autoref{fig:Duan}\textcolor{blue}{(a)} shows that the violation of the criteria occurs at lower two-photon detuning $\delta $, corresponding to regions of lower gain. 
As $\delta $ increases, the 4WM process moves into higher gain regimes, diminishing the violation of the criteria to the point where the entanglement signature is lost.
Figure \ref{fig:Duan}\textcolor{blue}{(b)},  shows the DGCZ criterion obtained from the theoretical model given in Eq. (\ref{eq:d_field}).
As in the experiment, the model predicts a reduced violation of the criterion as the two-photon detuning $\delta$ increases, however, not as steep as in the experimental case.

\MM{Figures \ref{fig:Duan}\textcolor{blue}{(c)} and \ref{fig:Duan}\textcolor{blue}{(d)} present the results in the sideband basis, corresponding to the subspace made out of pairs of modes shown in Fig. \ref{fig:Gain_amp}\textcolor{blue}{(a)}, specifically pairs $[\Omega _{pr}, -\Omega _{cj}]$ and $[-\Omega _{pr}, \Omega _{cj}]$}. At smaller $\delta $, both bipartitions show a violation of the criteria. However, as the detuning moves closer to the maximum gain, the mode $[\Omega _{pr}, -\Omega _{cj}]$ loses the violation of the criteria for frequencies $\delta/2\pi \geq 68$~MHz while the DGCZ test in the SA basis still presents some entanglement. On the other hand,  the mode $[-\Omega _{pr}, \Omega _{cj}]$ presents a more robust violation of the criteria than the SA basis over the values of $\delta$ in the range $\delta/2\pi \geq 68$~MHz.
Figure \ref{fig:Duan}\textcolor{blue}{(d)} presents the theoretical results that predict an asymmetry in the violation as $\delta$ approaches the region of maximum gain.
Although a steep change in the violation is observed in the experimental data for the range of 70~MHz $<\delta/2\pi <75$~MHz, the theoretical profile shows a monotonic growth. The causes of such effect are not known and require further investigation to determine their origin as well as any possible limitations of the model given in Ref. \cite{glorieux2010double}.


The asymmetry between the upper and lower sidebands can be better understood if we look at another entanglement criteria, in particular by mapping the separability test provided by the positivity under partial transposition (PPT) of the density matrix into the covariance matrix \cite{simon2000peres}. 
\MM{In the CV regime, the partial transposition corresponds to a mirror reflection of one quadrature of a subset of the system, given by the modes $k$. In particular, this transformation reverses the sign of the momentum operator sign for these modes ($\q _k\rightarrow - \q _k$).}
\MM{For separable states, the partially transposed covariance matrix, $\mathbb{V}_{PPT}$, remains physical, corresponding to a valid covariance matrix. Conversely, if $\mathbb{V}_{PPT}$ is unphysical, the bipartitions between the modes $k$ and the rest of the system are entangled. The physicality of $\mathbb{V}_{PPT}$ is determined by computing its smallest symplectic eigenvalue; the unphysicality is indicated when this eigenvalue falls below the threshold for separable states, which is normalized to one in this paper.
For two-mode systems, this provides a necessary and sufficient test for bipartite entanglement. For the PPT criterion, an entanglement witness comes from the smallest symplectic eigenvalue of the partially transposed matrix, which should be greater than unity for physical covariance matrices.}

\begin{figure}[htb]
    \centering
    \captionsetup{justification=raggedright, width=0.46\textwidth}
    \begin{overpic}[width=0.49\linewidth]{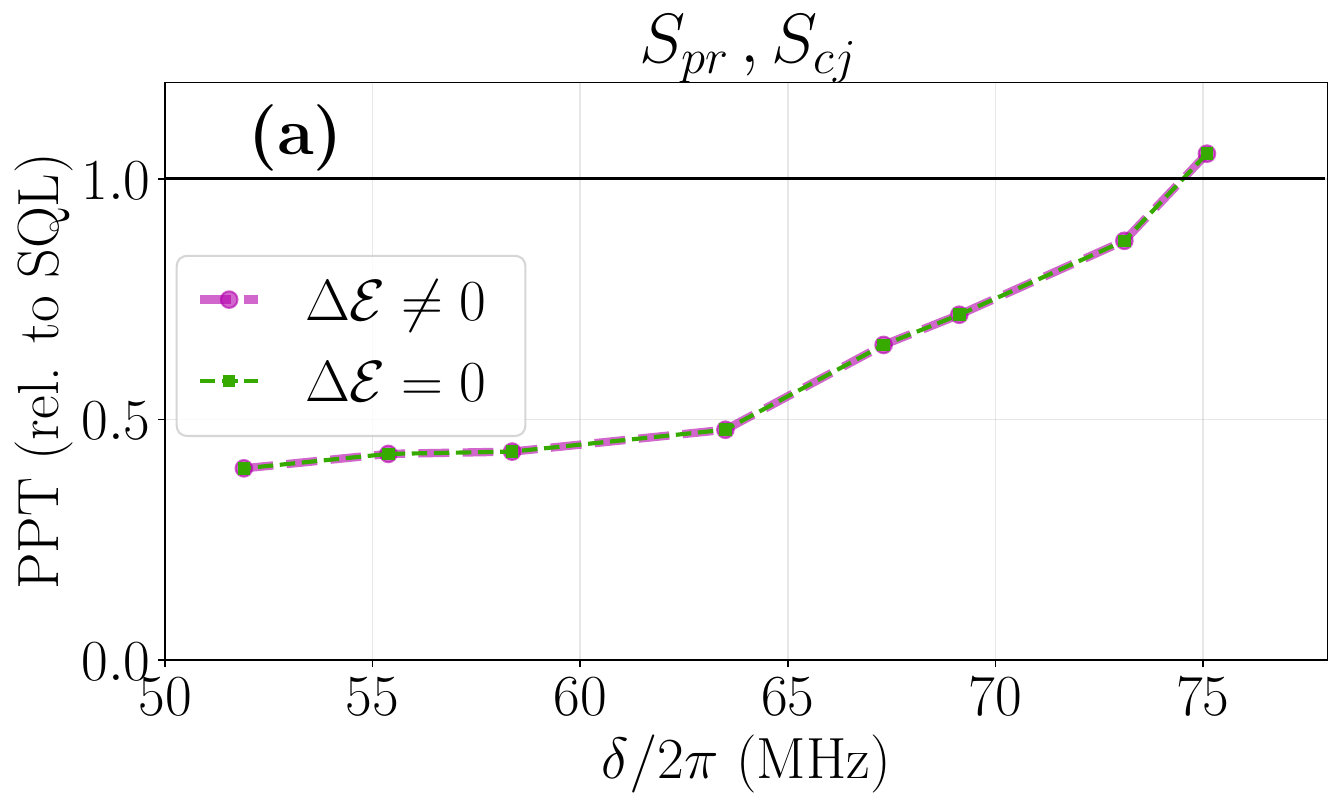}
    \end{overpic}%
    \begin{overpic}[width=0.49\linewidth]{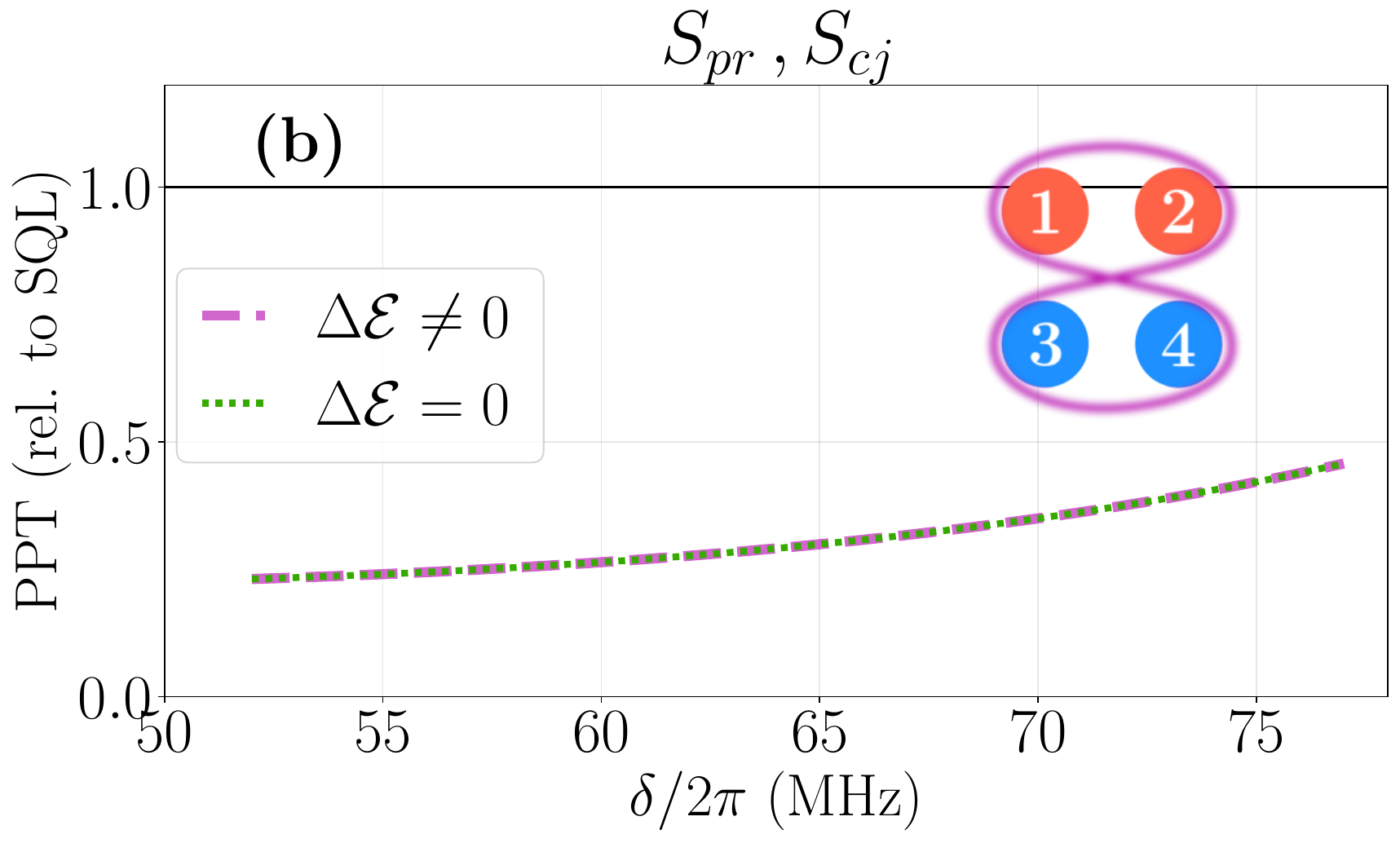}
    \end{overpic}
    \begin{overpic}[width=0.49\linewidth]{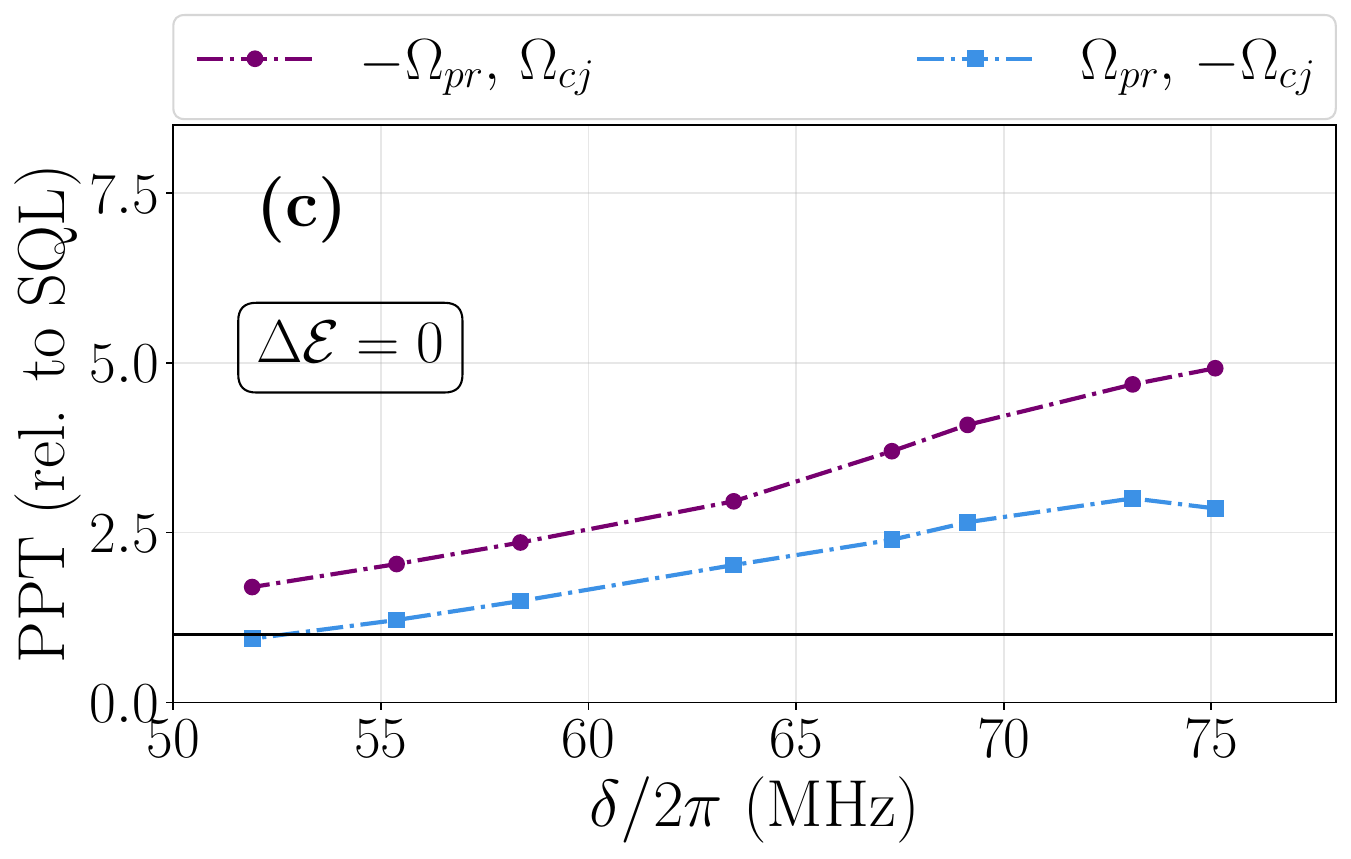}
    \end{overpic}%
    \begin{overpic}[width=0.49\linewidth]{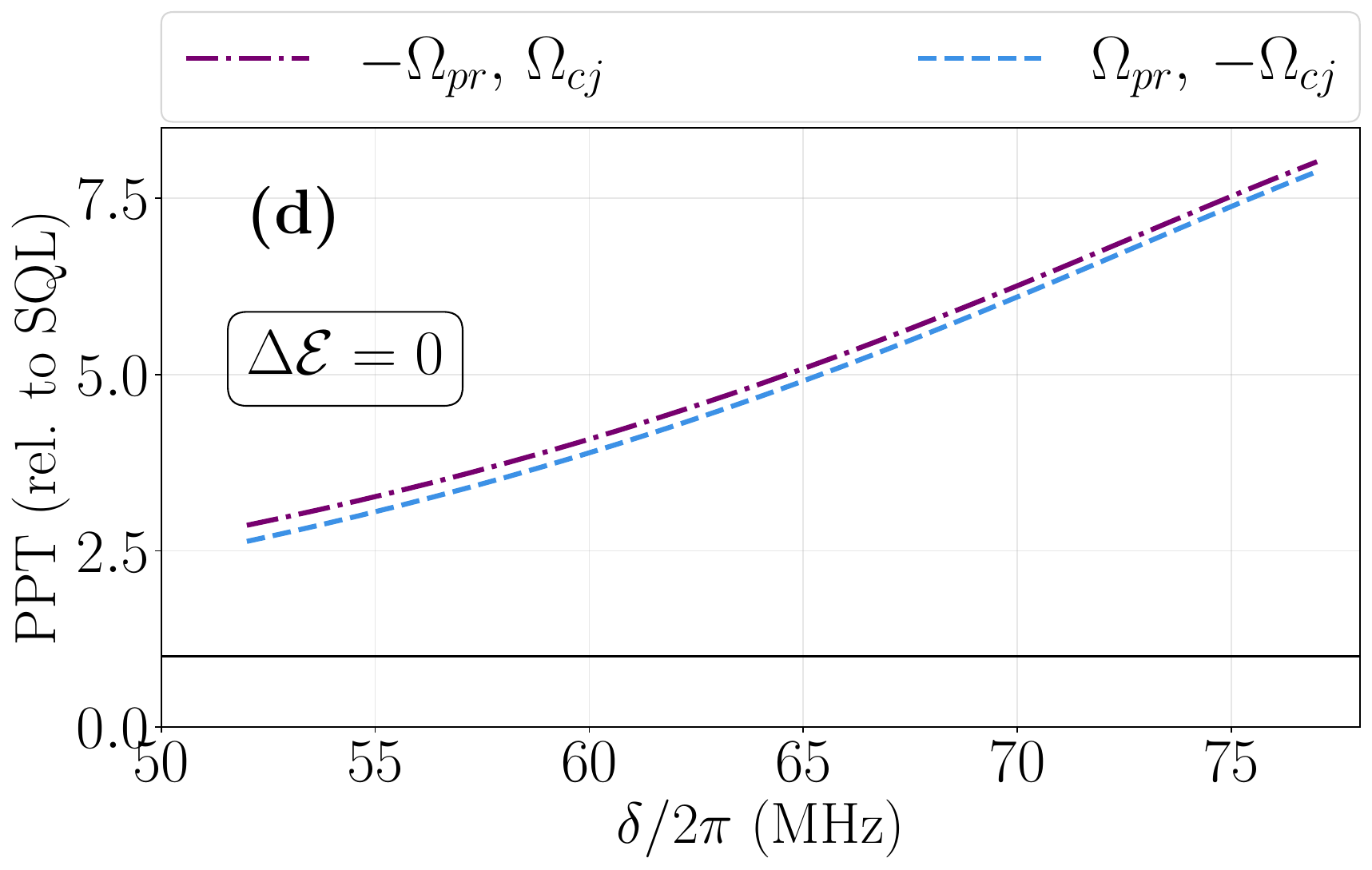}
    \end{overpic}
    \begin{overpic}[width=0.49\linewidth]{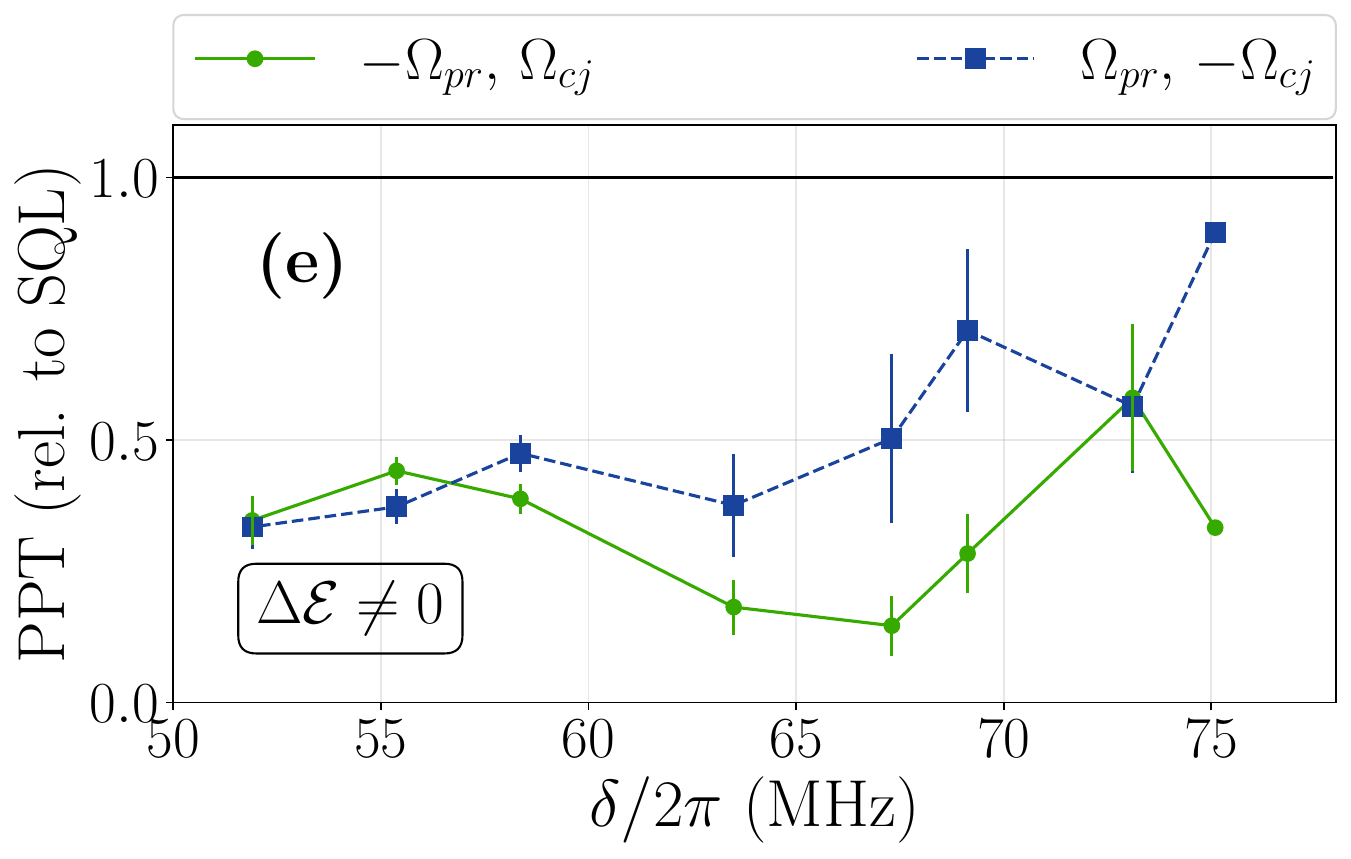}
    \end{overpic}%
    \begin{overpic}[width=0.49\linewidth]{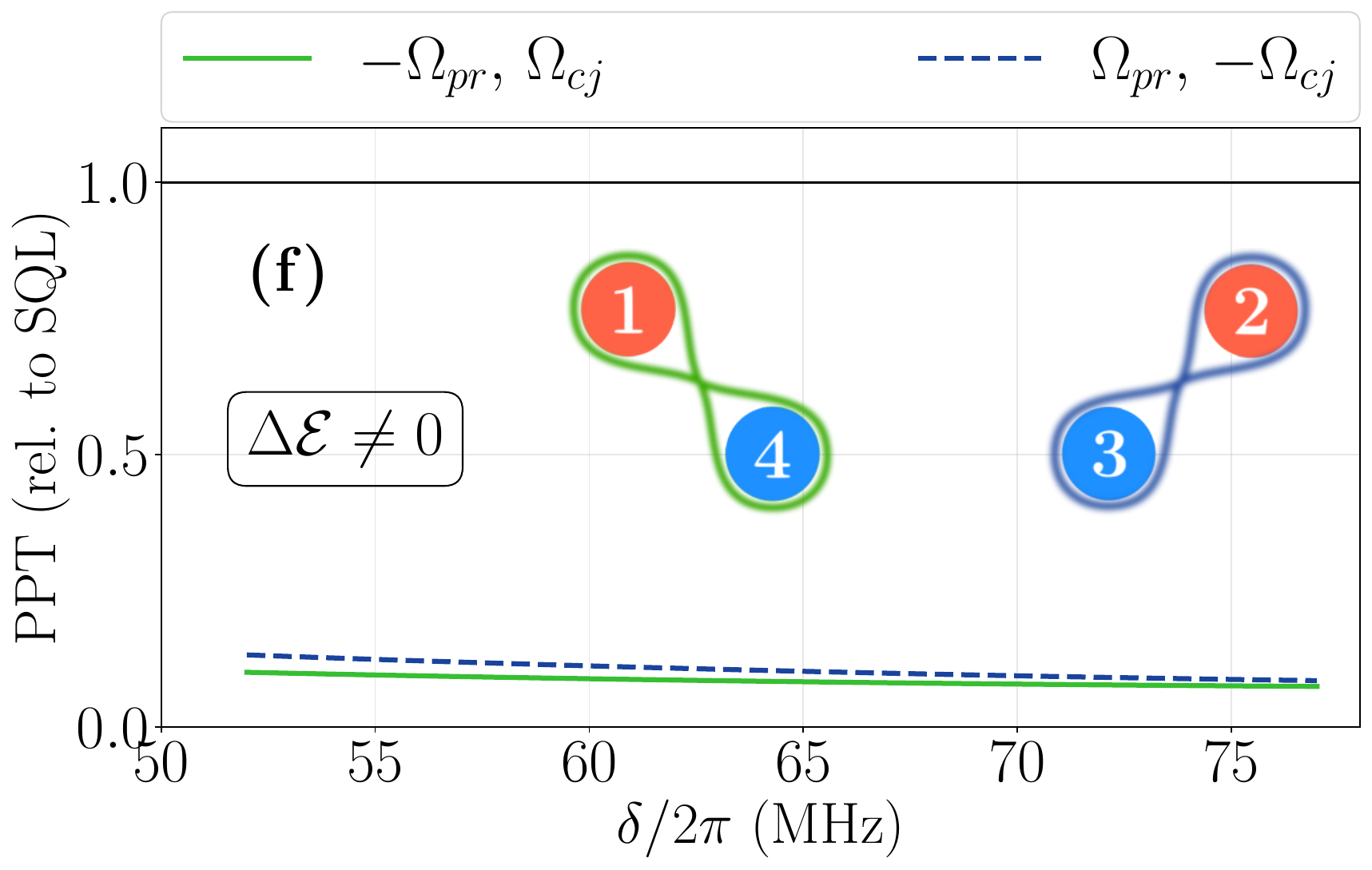}
    \end{overpic}
    \caption{PPT criteria: experimental and theoretical results in the SA basis, \textbf{(a)} and \textbf{(b)} respectively. In the sideband basis, \textbf{(c)} and \textbf{(d)} corresponds to the experimental and theoretical results considering the case when there is no access to the imbalance between the sideband modes; \textbf{(e)} and \textbf{(f)} corresponds to the experimental and theoretical results under the resonator-assisted detection. \label{fig:PPT}}
\end{figure}


\MM{\autoref{fig:PPT} depicts the PPT criteria for the subspace of pairs of sidebands, as previously described in Fig. \ref{fig:Duan}. Figs. \ref{fig:PPT}\textcolor{blue}{(a)} and \ref{fig:PPT}\textcolor{blue}{(b)} show the results in the SA basis, while Figs. \ref{fig:PPT}\textcolor{blue}{(c)}, \ref{fig:PPT}\textcolor{blue}{(d)}, \ref{fig:PPT}\textcolor{blue}{(e)}, and \ref{fig:PPT}\textcolor{blue}{(f)} show the results for the sideband basis. The left column corresponds to experimental results while the right column depicts theoretical results.} \HM{We also studied two distinct situations for the reconstructed covariance matrix to compare methods of detection. 
In the first scenario, we assume $\Delta \mathcal{E}= 0$, in order to reproduce the results as they would be if obtained by homodyne detection, which is unable to retrieve these correlations. In the second scenario, we consider the actual energy imbalance between the sidebands of the generated beams, such that $\Delta \mathcal{E}\neq 0$, which corresponds to the resonator detection method that successfully retrieves this information.}

\MM{In the SA basis in Figs.\ref{fig:PPT}\textcolor{blue}{(a)} and \ref{fig:PPT}\textcolor{blue}{(b)}, the symplectic eigenvalues increase as the probe approaches the maximum gain, eventually loosing the entanglement. Notably, the analysis of correlations on} \HM{the SA basis }\MM{is unable to distinguish between scenarios where $\Delta\mathcal{E}=0$ and $\Delta\mathcal{E}\neq 0$. }\HM{In contrast, this} \MM{ analysis can be extended to the sideband basis. In the case where $\Delta\mathcal{E}= 0$, as shown in Figs.  \ref{fig:PPT}\textcolor{blue}{(c)} and \ref{fig:PPT}\textcolor{blue}{(d)}} \HM{which emulates the state reconstruction by homodyne detection,} \MM{no violation of the PPT criteria is observed. However, when the detection method is able to retrieve the energy imbalance between the generated sideband modes, i.e., when $\Delta\mathcal{E}\neq 0$ as shown in Figs. \ref{fig:PPT}\textcolor{blue}{(e)} and \ref{fig:PPT}\textcolor{blue}{(f)}, the entanglement is preserved in both bipartitions.} 
\HM{In addition to preserving the entaglement in Fig.~\ref{fig:PPT}\textcolor{blue}{(e)}, notice that the symplectic eigenvalues show an asymmetry as the two-photon detuning increases and approaches the region of maximum gain.}
\MM{This asymmetry favors the entanglement in the pair involving the upper sideband of the probe and the lower sideband of the conjugate, similar to what was observed with the DGCZ criterion (Fig. \ref{fig:Duan}\textcolor{blue}{(c)}). 
\HM{ Therefore, the resonator detection  shows the hidden entanglement between sideband modes,  that cannot be detected otherwise. Unlike the demonstration in a previous work~\cite{Assumption2020}, where the asymmetry was introduce artificially by a cavity, here we present a system that intrinsically presents such asymmetry and the hidden entanglement.}
The theoretical model gives some qualitative agreement, as can be observed by the reduction in the violation for the SA basis, the robustness of the entanglement in the sideband basis, and the consistency across the two scenarios. Yet, it fails in explaining the asymmetry in the sideband basis, or the disentanglement in the SA basis.}


\HM{Such} robustness of the PPT entanglement criteria under the variation of the two-photon detuning $\delta $, in opposition to the DGCZ criteria, is a consequence of the fact that the DGCZ criterion deals only with the $\hat p$ and $\hat q$ quadratures, ignoring the correlations and imbalances of their variances, while the PPT criterion works with the whole covariance matrix, being an iff criterion for entanglement for two mode Gaussian states. We are thus led to the conclusion that some information is present in the terms $\langle \hat p_i \hat q_j \rangle$. These terms could be a result of self and cross phase rotations involving the generated fields and the pump, which would redistribute correlations of the kind $\langle \hat p_i \hat p_j \rangle$ and $\langle \hat q_i \hat q_j \rangle$ over the covariance matrix.


The conclusions coming from the detailed characterization of the sidebands of the fields produced by parametric amplification in the 4WM process in rubidium atoms point to the limitations of entanglement characterization by the usual DGCZ criteria, which only considers the variances in the symmetric/antisymetric basis. If we look at the detailed structure of the sidebands (Fig. \ref{fig:Duan}), we notice that entanglement is still observed for a pair of modes when the gain is higher. \MM{Considering the generated modes as Gaussian states, the PPT criterion can be used as a necessary and sufficient criterion, revealing a stronger violation of the criteria when studied in the sideband basis compared to the standard measurements that occur in the SA basis.}

These differences come from two origins. Phase rotation of the sideband modes with respect to the carrier, which could be produced by nonlinear processes, leads to information being redistributed among quadrature correlations between both beams, which represents an information loss in the na\"ive DGCZ analysis that is recovered in the PPT criterion.
Moreover, relevant information is lost if we ignore the terms that are unveiled only in the resonator-assisted detection, such as the power imbalance between sidebands (Fig. \ref{fig:delta_param}), \MM{that is impossible to retrieve using homodyne detection \cite{barbosa2013beyond}. }
These terms are essential for the complete reconstruction of the covariance matrix in the sideband basis, where robust entanglement is present between specific sideband pairs (Figs. \ref{fig:PPT}\textcolor{blue}{(e)} and \ref{fig:PPT}\textcolor{blue}{(f)}).

The theoretical model given in Ref. \cite{glorieux2010double} shows a qualitative agreement with the reconstructed covariance matrix (see Supplementary Material) and provides an explanation for many of the observed trends in the pairwise entanglement production involving the sidebands. However, it is not able to provide a quantitative reproduction of the exact behavior of the entanglement witnesses. For the DGCZ criterion, the inability of the theory to observe the loss of entanglement can be justified by phase rotations that the model does not take into account, thus requiring a more detailed analysis, which is outside the scope of our current work and will be further addressed in the future.



\MM{Finally, we note that the full characterization of quantum correlations in the 4WM process paves the way for generating multi-mode entangled states within the narrow gain bandwidth of these atomic systems, while also providing insight into the intricate inner spectral structure of quantum correlations in the parametric amplification process using atoms.}




This work was funded by Grants  No. 2015/18834-0, 2017/27216-4, 2018/03155-9 and 2019/12840-0  (S\~ao  Paulo  Research  Foundation - FAPESP),  Grant No. N629091612184 (NICOP/ONR), and PHYS-1752938 (National Science Foundation - NSF grant). 


\nocite{*}

\bibliography{apssamp}

\end{document}